\documentclass[prl,aps,preprint]{revtex4}

\usepackage{amssymb}
\usepackage{graphicx}
\usepackage{amsmath}

\begin{document}

\title{Fermi surface evolution in an electron-doped high-temperature superconductor
Nd$_{2-x}$Ce$_x$CuO$_4$ revealed by Shubnikov-de Haas oscillations}
\author{T. Helm$^1$}
\author{M. V. Kartsovnik$^1$}
\author{M. Bartkowiak$^2$}
\author{N. Bittner$^1$}
\author{M. Lambacher$^1$}
\author{A. Erb$^1$}
\author{J. Wosnitza$^2$}
\author{R. Gross$^{1,3}$}

\affiliation{$^1$Walther-Mei{\ss}ner-Institut, Bayerische Akademie der
Wissenschaften, Walther-Mei{\ss}ner-Str. 8, D-85748 Garching, Germany}
\affiliation{$^2$ Hochfeld-Magnetlabor Dresden, Forschungszentrum
Dresden-Rossendorf, Bautzner Landstr. 400, D-01328 Dresden, Germany}
\affiliation{$^3$ Physik-Department, Technische Universit\"{a}t M\"{u}nchen,
James Franck Str., D-85748 Garching, Germany}

\begin{abstract}
We report on the direct probing of the Fermi surface in the bulk of the
electron-doped superconductor Nd$_{2-x}$Ce$_x$CuO$_4$ at different doping levels
by means of magnetoresistance quantum oscillations. Our data reveal a sharp
qualitative change in the Fermi surface topology, due to translational symmetry
breaking in the electronic system which occurs at a critical doping level
significantly exceeding the optimal doping. This result implies that
the $(\pi/a,\pi/a)$ ordering, known to exist at low doping levels, survives up to
the overdoped superconducting regime.
\end{abstract}
\maketitle

The Fermi surface topology and its evolution with doping is one of the most
important issues related to the nature of charge carriers and various
competing ordering phenomena in the high-temperature superconducting
cuprates. The research has been pushed forward immensely by the recent
discovery of slow magnetic quantum oscillations in the hole-underdoped
superconductors YBa$_2$Cu$_3$O$_{6.5}$ \cite{doir07,jaud08,seba08} and
YBa$_2$Cu$_4$O$_8$ \cite{yell08,bang08}. While the general trend is to
associate these oscillations with small Fermi surface pockets, the latter
apparently contradict the discontinuous Fermi arcs scenario derived from
angle-resolved photoemission (ARPES) experiments \cite{dama03} and their
origin is currently a matter of hot debate
\cite{doir07,seba08,bang08,mill07,lebo07,podo08,lee08,chen08,kaul08,dimo08,alex08,jia08}.
To solve the problem, it is pivotal to understand how the Fermi surface
develops with changing the concentration and even the sign of charge carriers.
However, direct probing of the Fermi surface by means of magnetic quantum
oscillations has been restricted so far to
hole-doped compounds with few selected doping levels, namely to the above mentioned
underdoped yttrium-barium cuprates characterized by highly ordered oxygen and to
cleanest samples of strongly overdoped Tl$_2$Ba$_2$CuO$_{6+\delta}$ \cite{vign08}.
Here we report the observation of quantum oscillations in the interlayer
magnetoresistance of the electron-doped superconductor Nd$_{2-x}$Ce$_x$CuO$_4$
(NCCO) at the optimal, $x=0.15$, as well as overdoped, $x=0.16$ and 0.17,
compositions.
The data obtained provide direct evidence for a well-defined closed Fermi
surface (rather than Fermi arcs) and clearly reveal its evolution with doping
level. In particular, a dramatic change in the oscillation spectrum observed at
increasing $x$ from 0.16 to 0.17 is indicative of a topological transformation of
the Fermi surface in the overdoped regime.

Compared to hole-doped cuprates, the electron-doped NCCO has a few advantages.
First, the superconducting state of NCCO is restricted to the narrow doping
interval $0.13\leq n \leq 0.18$ (Fig. 1), where $n$ is the number of doped
electrons per Cu ion and $n = x$ due to the well-defined valences of the
Nd$^{3+}$ and Ce$^{4+}$ ions.
Hence, the whole relevant doping range can be
controllably scanned by slightly varying $x$. Second, the upper critical field
of NCCO does nod exceed 10 T \cite{wang03} which is about an order of magnitude
lower than for hole-doped superconductors. Thus, for any doping level, it is
easy to suppress superconductivity at low temperatures by applying magnetic field
(Fig. 1). This excludes potential ambiguities and complications in the interpretation of
magnetic quantum oscillations associated with the vortex state of
superconductors \cite{alex08,dimo08,jia08,mani01}. Finally, the Fermi surface of
NCCO is expected to be relatively simple: the material only contains a
single conducting CuO$_2$ layer per unit cell, avoiding ambiguities with any
bilayer potential \cite{bang08,podo08}.

Single crystals of NCCO were grown in an Ar/O$_2$
atmosphere using the travelling solvent floating zone method and annealed
in pure argon at $950^{\circ}$~C for 20 h in order to remove interstitial
oxygen and strain in the crystal lattice. The high quality of the samples was
ensured by structural analysis as well as by magnetic and resistive
measurements. The superconducting transition temperatures (taken as the midpoint
of the transition in magnetic susceptibility) were $T_c = 23.5, 19.2$, and 5.9 K
for the crystals with $x=0.15$, 0.16, and 0.17, respectively, in agreement with
detailed phase diagram studies \cite{lamb08}. The transition widths were $< 1$ K,
2.5 K, and 3.5 K for $x = 0.15$, 0.16, and 0.17, respectively, thus, ensuring homogeneity
of Ce doping within $\sim 0.25\%$. Measurements of the interlayer resistance were
performed using an a.c. current of 0.4-1.0 mA at a frequency
of 67 kHz. The magnetic field was applied perpendicular to the CuO$_2$ planes
($\mathbf{B} \parallel c$-axis), using a 70 T pulse magnet with a pulse duration of
150 ms at the Hochfeld-Magnetlabor Dresden. While the data presented here were
taken at the decaying part of the pulse, they were reproduced, although with a
higher noise level, at the rising part.

Fig. 2(a) shows the oscillatory component of the interlayer resistivity
$\rho_{\mathrm{osc}}$ of the optimally doped, $x = 0.15$, sample, obtained from
the raw data (Fig. 1) after subtracting the monotonic background.
The oscillations are periodic in $1/B$ [Fig. 2(b)] with $T$-independent
positions of their maxima and minima, whereas their amplitude $A$ gradually
increases upon cooling. Such behavior is a prominent characteristic of
the Shubnikov-de Haas (SdH) effect, originating from the
Landau quantization of the quasiparticle spectrum \cite{abri88}. It, thus,
provides clear evidence for a well-defined closed Fermi surface, made up of
fermionic charge carriers. The oscillation
frequency, $F_{0.15} = (290\pm 10)$ T, yields the area of the extremal Fermi
surface cross-section, $S_{0.15} = 2\pi eF_{0.15}/\hbar = (2.75 \pm 0.10)$
nm$^{-2}$.
This area amounts to only 1.1\% of the two-dimensional Brillouin zone,
$S_{\mathrm{BZ}} = (2\pi /a)^2 = 253$ nm$^{-2}$, where $a = 3.95$ {\AA} is the
in-plane lattice constant of NCCO.

The whole data set in Fig. 2(a) can be described by the standard Lifshitz-Kosevich
(LK) formula for magnetic quantum oscillations \cite{abri88}:
\begin{equation}
\rho_{\mathrm{osc}} \propto
B^{1/2}R_TR_{\mathrm{D}}\sin \left(2\pi F/B + \gamma \right),
\label{LK}
\end{equation}
where $R_T=\alpha T/B\sinh(\alpha m_cT/B)$ and
$R_{\mathrm{D}}= \exp(-\alpha m_cT_{\mathrm{D}}/B)$
are, respectively, the temperature and scattering damping factors,
$\alpha=14.69$ T/K, $m_c$ is the effective cyclotron mass in units of the free
electron mass, $T_{\mathrm{D}}$ is the Dingle temperature determined by
scattering, and $\gamma$ is the temperature and field independent Onsager phase.
Figs. 2(c),(d) show examples of the temperature and field dependence of the
oscillation amplitude yielding, respectively, the cyclotron mass
$m_c = 0.6 \pm 0.05$ and Dingle temperature $T_{\mathrm{D}} \approx 15$ K.
The latter provides an estimate for the scattering time
$\tau \simeq \hbar/2\pi k_{\mathrm{B}}T_{\mathrm{D}}
\approx 0.8\times 10^{-13}$ s. This corresponds to a mean free path, averaged
over the cyclotron orbit, of
$\ell \sim \hbar(S/\pi )^{1/2}\tau/m_c \approx 14$ nm.
Using the above mentioned values for $F$, $m_c$, and $T_{\mathrm{D}}$, we
obtain a remarkably good fit of the experimental data to Eq. (\ref{LK})
in the entire temperature and field range studied, as demonstrated by dotted
lines in Fig. 2(a).

The moderately overdoped sample ($x = 0.16$) shows similar quantum oscillations,
however, with only about half the amplitude [Fig. 3(a)]. The oscillation frequency,
$F_{0.16} = (280 \pm 15)$ T, is almost the same as for optimal doping. Whereas
the overall behavior is similar for the $x = 0.15$ and $x = 0.16$ samples, a
drastic change is observed on further increasing the doping level to $x = 0.17$.
The slow SdH oscillations vanish and, instead, fast oscillations, also periodic
in $1/B$, emerge at fields above 60 T [Fig. 3(b)]. Their frequency,
$F_{0.17} = (10.7 \pm 0.4)\times 10^3$ T, corresponds to a large cyclotron orbit on
the Fermi surface enclosing the area $S_{0.17} = (102 \pm 4)$ nm$^{-2}$ in
$\mathbf{k}$-space or, equivalently, $(0.405 \pm 0.015)S_{\mathrm{BZ}}$.

The markedly different $F$ obtained for the strongly overdoped ($x = 0.17$) sample
on the one hand and for the optimally doped ($x = 0.15$) and slightly overdoped
($x = 0.16$) samples on the other hand becomes apparent in Fig. 3(c). For $x = 0.17$,
the interpretation of our data is straightforward: the size of the cyclotron orbit
is fully consistent with the results of band-structure calculations \cite{mass89}
and ARPES \cite{mats07,armi02} suggesting a single Fermi cylinder centered at the
corner of the Brillouin zone, as shown in Fig. 4(a). For $x = 0.17$, we expect
$S_{0.17} = 0.415S_{\mathrm{BZ}}$ in perfect agreement with our experimental result.
In contrast, the slow oscillations observed at the lower doping levels
reveal a very small Fermi surface, indicating a qualitative change in its
topology. This suggests that we do not deal with a conventional gradual evolution
of the Fermi surface with doping, but rather with a reconstruction due to a broken
symmetry. This transformation can be explained by assuming that the commensurate
density-wave superstructure, appearing in the electronic system of undoped and
underdoped NCCO \cite{armi02}, survives in the optimally doped and even slightly
overdoped regime \cite{mark07a}. The ordering potential splits the original conduction
band described by the dispersion \cite{ande95,lin07}:
\begin{eqnarray}
\varepsilon_{\mathbf{k}} & = & -2t(\cos ak_x+\cos ak_y)+4t'\cos ak_x \cos ak_y -
\nonumber \\
&&  2t'' (\cos 2ak_x + \cos 2ak_y) + \mu ,
\label{tbd}
\end{eqnarray}
where $\mu$ is the chemical potential determined by the doping level, into two bands
\begin{equation}
\varepsilon_{\mathbf{k}}^{\pm} =
\frac{\varepsilon_{\mathbf{k}}+\varepsilon_{\mathbf{k}+ \mathbf{Q}}}{2}
\pm  \sqrt{\left(
\frac{\varepsilon_{\mathbf{k}}-\varepsilon_{\mathbf{k}+ \mathbf{Q}}}{2}
\right)^2 + \Delta ^2}.
\label{twoband}
\end{equation}
Here, $\mathbf{Q} = (\pi/a, \pi/a)$ is the superstructure wave vector and $\Delta$
is the energy gap between the lower and upper bands determined by the strength of
the superstructure potential. As a result, the original large Fermi surface
is folded and split, forming one electron and two small hole pockets in the new
Brillouin zone as shown in Fig. 4(b).
Assuming that the slow SdH oscillations originate from the
hole pockets and using literature values for the effective overlap integrals
\cite{ande95,lin07}: $t  = 0.38$ eV, $t' = 0.32t$, and $t''   = 0.5 t'$  one can
apply Eqs. (\ref{tbd}) and (\ref{twoband}) to fit the size of the pockets to that
obtained from the oscillation frequency, with the gap $\Delta$ as the fitting
parameter. This yields $\Delta  = 64$ meV ($x = 0.15$) and 36 meV ($x = 0.16$).
Alternatively, one could ascribe the oscillations to the remaining electron
pocket by assuming a much larger gap which would completely destroy the
hole pockets. However, this would imply an unreasonably large gap,
$\Delta = 0.64$ eV, comparable to that in the undoped mother compound \cite{armi02}.
Moreover, in this case the carrier concentration
$n = 4S_{0.15}/S_{\mathrm{BZ}} \approx 0.045$ electrons per copper site would
be totally inconsistent with the nominal value, $n = 0.15$, for optimal doping.
Hence, it is most likely that the SdH oscillations originate from the small
hole pockets of a reconstructed Fermi surface.

It is tempting to assign the observed transition from a Fermi surface made of
small pockets at $x \leq 0.16$ to a large Fermi surface at $x = 0.17$ to a
topological change at a critical doping level. Although this scenario is
likely, at present we cannot unequivocally rule out the Fermi-surface
reconstruction at $x = 0.17$ due to a possible magnetic breakdown \cite{abri88},
allowing charge carriers to tunnel through a small gap which separates
different parts of the Fermi surface in $\mathbf{k}$-space. In our case, small orbits
on the reconstructed Fermi surface [Fig. 4(b)] would be considerably suppressed,
giving way to the large orbit shown in Fig. 4(a) at a field exceeding the
breakdown field
$B_{\mathrm{MB}} \simeq 4\Delta ^2m_c/\hbar e \varepsilon_{\mathrm{F}}$.
Here, $\varepsilon_{\mathrm{F}}$ is the Fermi energy taken from the top of
the conduction band. While more detailed data is necessary for a quantitative
analysis, we already can estimate the upper limit for the superstructure gap
at $x = 0.17$ to $\Delta_{0.17} \leq 8$ meV by setting
$B_{\mathrm{MB},0.17}\leq 30$ T.

The combined data of our study and recent experiments
\cite{doir07,jaud08,seba08,yell08,bang08} suggests that in both hole- and
electron-doped superconducting cuprates the Fermi surface undergoes a
reconstruction below a certain doping level.
Whereas for the hole-doped cuprates no conclusive data on the
evolution of the Fermi surface around optimal doping is available so far,
our study of the electron-doped NCCO clearly shows that the reconstructed
Fermi surface is present at optimum doping and even persists into the
overdoped regime. This conclusion is consistent with the inplane
magnetotransport studies \cite{lin07,lamb08,daga04} suggesting the existence
of two types of carriers in electron-overdoped superconductors and pointing
to the dominant role of hole-like carriers in conductivity.

On the other hand, our results apparently contradict ARPES and inelastic
neutron scattering data on NCCO. In the ARPES experiments \cite{mats07,armi02},
no small hole-like Fermi pockets were observed. Instead, a finite spectral
weight at the intersections of the large Fermi surface with the reduced
Brillouin zone boundary [dashed line in Fig. 4(b)] was reported for optimally
and overdoped  samples, suggesting no long-range ordering. The recent neutron
scattering measurements \cite{moto07} have revealed an antiferromagnetic
correlation length
$\xi _{\mathrm{AF}} \leq 5$ nm in optimally doped NCCO. However, our
observation of the slow SdH oscillations in samples with $x = 0.15$ and
0.16 starting from $B \approx  30$ T implies the correlation length being
at least of the order of the size of the coherent cyclotron orbit,
$\sim 2p_{\mathrm{F}}/eB \approx 40$ nm, taking the average
in-plane Fermi momentum
$p_{\mathrm{F}} = \hbar(S_{0.15}/\pi)^{1/2} \approx  10^{-25}$ kg m s$^{-2}$.
A similar problem holds for the hole-underdoped YBa$_2$Cu$_3$O$_{6.5}$ and
YBa$_2$Cu$_4$O$_8$: the quantum oscillations point to a long-range ordering,
which has not been revealed by any other experiment so far. While further
studies are necessary for establishing the exact origin of the $(\pi/a,\pi/a)$
ordering revealed by the quantum oscillations, a plausible scenario
is a field-induced antiferromagnetic ordering \cite{chen08}. Indeed, indications of enhanced
antiferromagnetic correlations in a magnetic field have been reported both for hole-doped
\cite{lake02,haug09} and for electron-doped \cite{mats03} cuprate superconductors.

We thank W. Biberacher for valuable discussions, and the staff of
Walther-Mei{\ss}ner-Institut, the Hochfeld-Magnetlabor Dresden-Rossendorf, and the
Kristalllabor of TUM for technical assistance. The work was supported by
EuroMagNET under the EU contract RII3-CT-2004-506239, by the German Science
Foundation via the Research Unit FOR 538, and by the German Excellence Initiative
via the Nanosystems Initiative Munich.

\newpage
\begin{figure}
	\centering
		\includegraphics[width=0.9\textwidth]{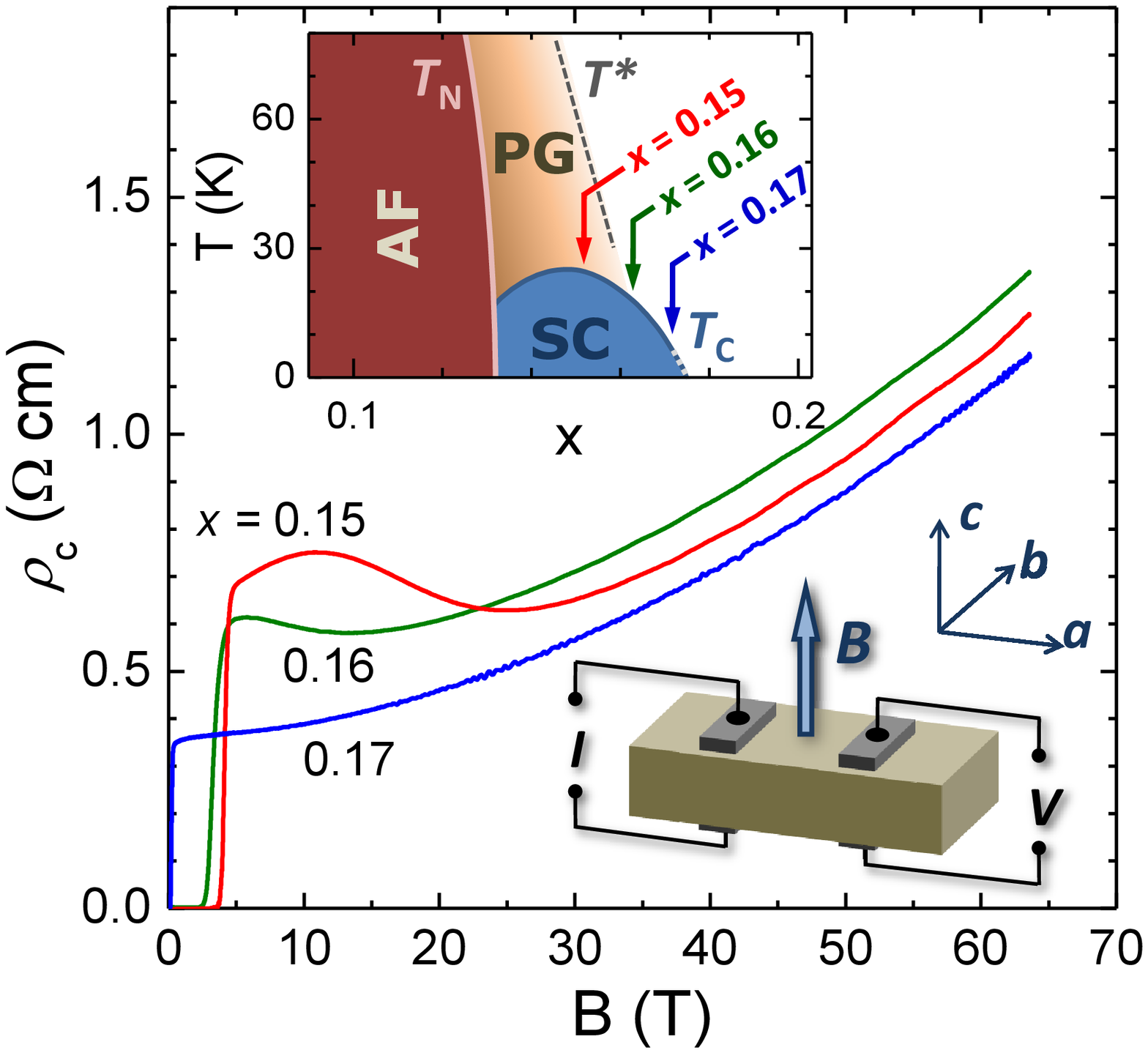}
	\caption{ (Color online). $c$-axis resistivity $\rho_c$
of NCCO plotted vs. magnetic field
applied perpendicular to the CuO$_2$ planes at $T = 4$ K for
different doping levels $x$. The upper inset
shows schematically the currently accepted phase diagram of NCCO with the
superconducting (SC), antiferromagnetic (AF), and pseudogap (PG) regions.
The arrows mark the compositions studied in this work. The lower inset
illustrates the geometry of the experiment. 
}
	\label{fig:fig1}
\end{figure}
\newpage
\begin{figure}
	\centering
		\includegraphics[width=0.9\textwidth]{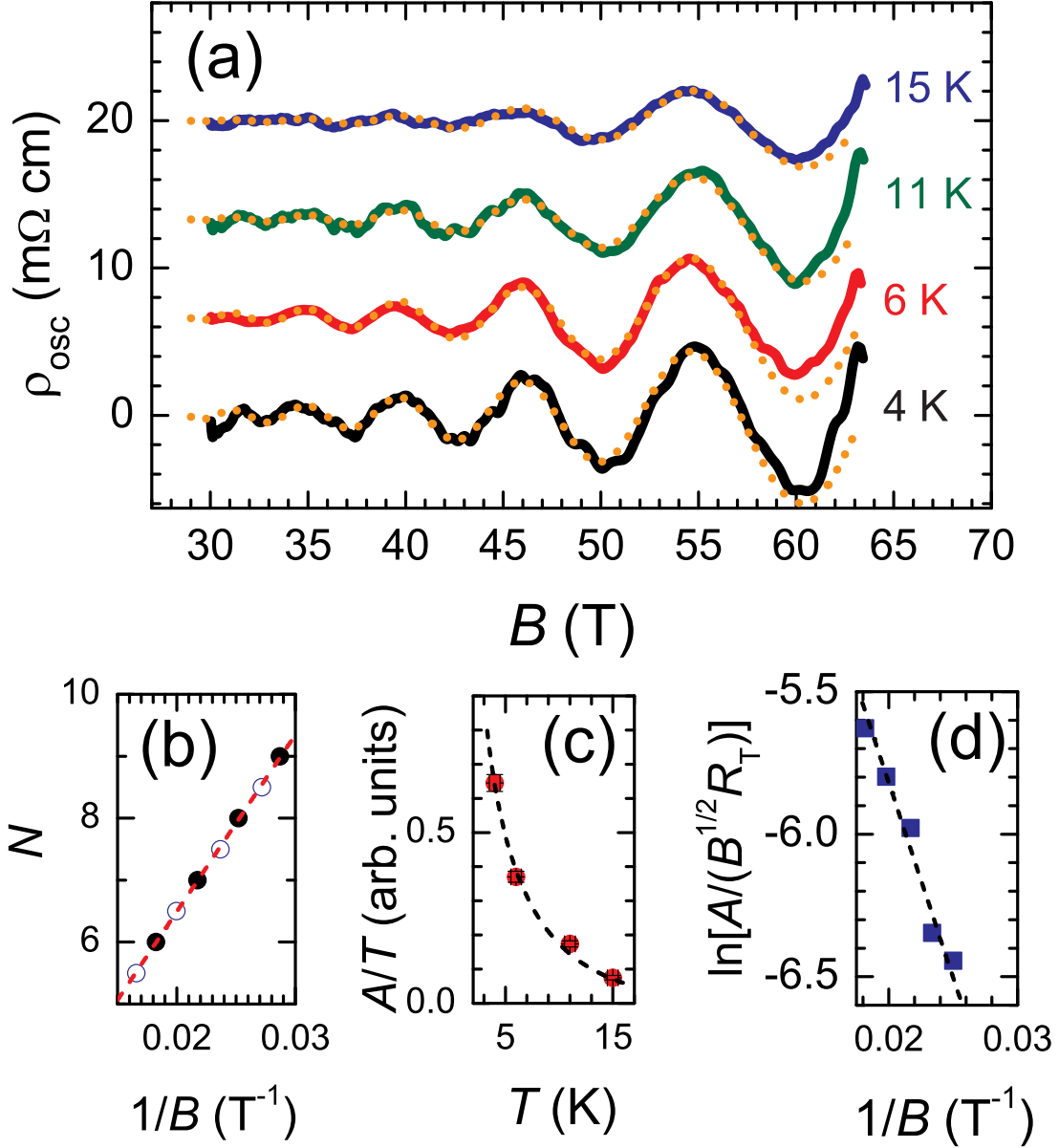}
	\caption{ (Color online). (a) Oscillatory part of the interlayer resistivity 
of optimally doped NCCO as a function of a magnetic field 
($\mathbf{B} \parallel c$-axis) at different $T$. The dotted lines are LK fits 
to the data using the values $F$, $m_c$, and $T_{\mathrm{D}}$ obtained from 
plots in (b), (c), and (d), respectively. (b) Positions of the local maxima 
(solid circles) and minima (open circles) of $\rho_{\mathrm{osc}}$ on an 
inverse field scale. A linear fit (dotted line) yields $F = 290$ T. 
(c) Temperature dependence of the oscillation amplitude at $B = 55$ T. 
The dashed line is the LK temperature dependence with $m_c = 0.6m_e$. 
(d) Dingle plot of the oscillation amplitude at $T = 4.0$ K, yielding 
the Dingle temperature $T_{\mathrm{D}} = 15$ K. 
}
	\label{fig:fig2}
\end{figure}
\newpage
\begin{figure}
	\centering
		\includegraphics[width=0.9\textwidth]{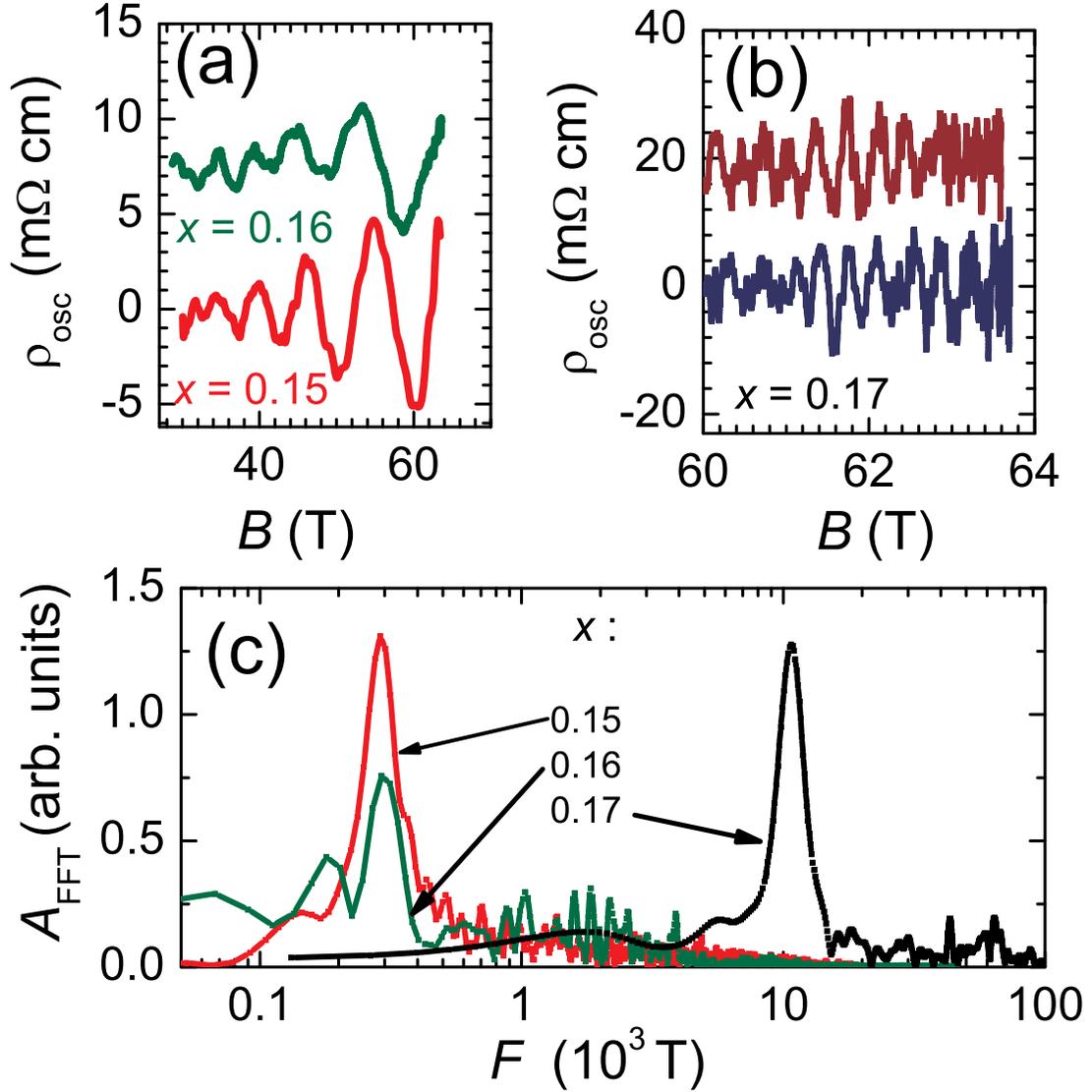}
	\caption{ (Color online). Doping dependence of SdH oscillations in NCCO. 
(a) Slow oscillations in the optimally doped and slightly overdoped 
($x = 0.16$) samples at $T = 4.0$ K. 
(b) Fast oscillations in the sample with $x = 0.17$; $T = 3.5$ K. 
Data from two different field pulses are shown. (c) Corresponding fast 
Fourier transform spectra of the oscillatory resistivity. For 
$x= 0.17$, the spectrum corresponds to an average of the two data sets 
shown in (b).
}
	\label{fig:fig3}
\end{figure}
\newpage
\begin{figure}
	\centering
		\includegraphics[width=0.9\textwidth]{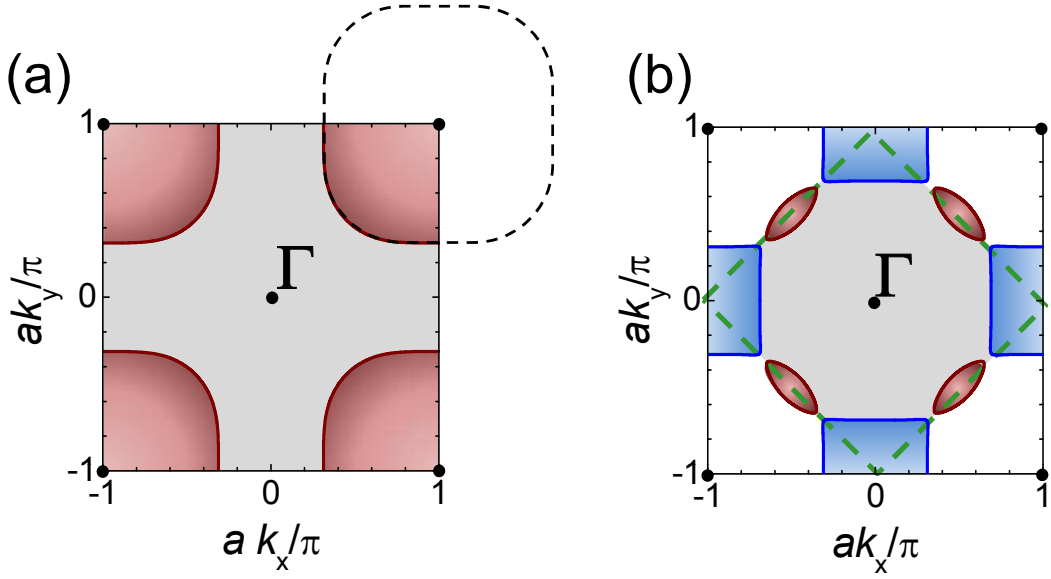}
	\caption{ (Color online). (a) Single-component Fermi surface with a 
cross-sectional area equal to $\approx 41\%$ of the first Brillouin 
zone corresponding to $x = 0.17$. The dashed line shows the large 
closed orbit, in the extended zone representation, responsible for the 
SdH oscillations. (b) Reconstructed Fermi surface comprising one electron 
and two hole pockets in the reduced Brillouin zone (dashed line). The hole 
pockets are responsible for the slow oscillations observed on the $x = 0.15$ 
and 0.16 crystals. 
}
	\label{fig:fig4}
\end{figure}

\end{document}